\documentclass[prl, twocolumn, showpacs, floatfix, amsmath, amssymb, superscriptaddress]{revtex4-1}
\usepackage{graphicx}

\usepackage{dcolumn}
\usepackage{hyperref}
\hypersetup{
    colorlinks=true,    
    linkcolor=blue,     
    citecolor=blue,     
    urlcolor=blue       
}
\usepackage{color}
\usepackage{url}
\begin{document}
\author{Chunyan Li}
\affiliation{Center for Advanced Quantum Studies, Department of Physics, Beijing Normal University, Beijing 100875}
\author{Haiwen Liu}
\email{haiwen.liu@bnu.edu.cn}
\affiliation{Center for Advanced Quantum Studies, Department of Physics, Beijing Normal University, Beijing 100875}
\affiliation{Interdisciplinary Center for Theoretical Physics and Information Sciences, Fudan University, Shanghai 200433, China}
\author{X. C. Xie}
\affiliation{Interdisciplinary Center for Theoretical Physics and Information Sciences, Fudan University, Shanghai 200433, China}
\affiliation{International Center for Quantum Materials, School of Physics, Peking University, Beijing 100871, China}
\affiliation{Hefei National Laboratory, Hefei 230088, China}

\title{The route of random process to ultraslow aging phenomena}

\begin{abstract}
Logarithmic aging phenomena are prevalent in various systems, including electronic materials and biological structures. This study utilizes a generalized continuous-time random walk (CTRW) framework to investigate the mechanisms behind the logarithmic aging phenomena. By incorporating non-Markovian jump processes with significant memory effects, we modify traditional diffusion models to exhibit logarithmic decay in both survival and return probabilities. In addition, we analyze the impact of aging on autocorrelation functions, illustrating how long-term memory behaviors affect the temporal evolution of physical properties. These results connect microscopic models to macroscopic manifestations in real-world systems, advancing the understanding of ultraslow dynamics in disordered systems.
\end{abstract}
\maketitle
\noindent

Logarithmic aging phenomena have been ubiquitously observed across a spectrum of complex systems, 
including conductance relaxations in Anderson insulators and electron glasses {\cite{Anderson_PhysRevLett.84.3402,Ovadyahu_PhysRevLett.92.066801}}, the evolution of frictional strength \cite{log_aging_nature2}, the dynamics of compaction in granular systems \cite{theory_aging_Shohat2023} , flux-creep dynamics in superconductors \cite{13prl_superconductor_PhysRevB.48.6477}, 
%anomalous electric-field effects in electron glasses ($1),
and logarithmic kinetics in various glassy systems 
\cite{13prl_glass_PhysRevLett.80.999,13prl_glass_doi:10.1073/pnas.1608057113,13prl_glass_doi:10.1126/science.287.5453.627}, as well as in the structural relaxation processes of proteins and DNA \cite{13prl_biology_PhysRevLett.125.058001,13prl_biology_PhysRevLett.88.158101,13prl_biology_nature}. 
A substantial body of theoretical and phenomenological research \cite{theory_aging_Shohat2023,theory_aging_PhysRevLett.99.226603,theory_aging_PhysRevLett.87.055502,theory_aging_PhysRevLett.84.5403,theory_aging_doi:10.1073/pnas.1120147109} supports these observations. 
Understanding the underlying mechanisms of ultraslow aging phenomena is crucial for predicting the long-term behavior of complex systems and improving the design of materials and devices with enhanced stability and performance.

\begin{figure*}[tbhp]
\centering
\includegraphics[width=17.8cm]{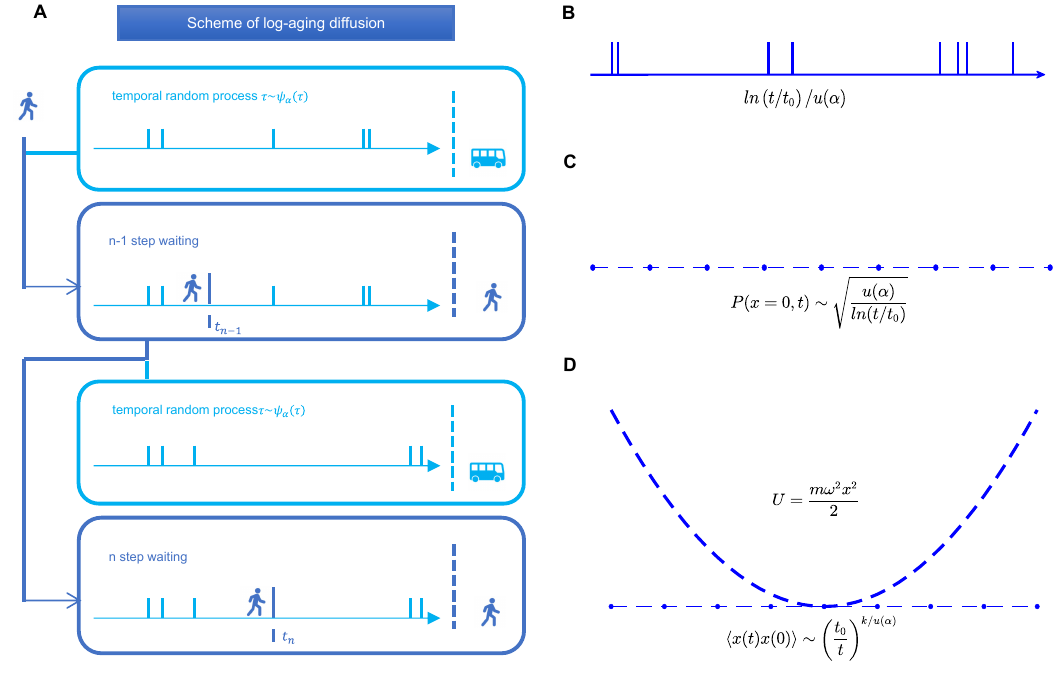}
\caption{Scheme and characteristics of log-aging diffusion.(\textbf{A}), We present a log-aging diffusion process, using a scenario where a passenger is waiting to board a bus. The intervals between bus departures at each stop are modeled by a long-tail distribution, expressed as $\psi_\alpha \sim \frac{1}{t^{1+\alpha}}$. Upon the passenger's arrival at the $n-1$ stop, they encounter a wait of $t_{n-1}$. Once the immediate bus arrives at the $n-1$ stop, the passenger boards swiftly and proceeds to the next stop $n$ with the travel time considerably shorter than the waiting time. Upon reaching stop $n$, the passenger waits for the next bus, timing the wait as $t_n$. (\textbf{B}), Displayed are the first nine states of a particle undergoing the diffusion process described in (A), parameterized by $\alpha=0.6$, $t_0=10^3$. We define $u\left(\alpha\right)=-\gamma-\partial \ln\Gamma\left(\alpha\right)/\partial \alpha$, where $\gamma$ is the Euler constant and $\Gamma$ is complete gamma function. (\textbf{C}), The return probability without external forces for the log-aging diffusion is computed as $P(x=0,t)$. (\textbf{D}), The positional autocorrelation function under the influence of an oscillator potential is given,  $C_x=\left<x(t)x(0)\right>$.}\label{fig1}
\end{figure*}

Previously, Lomholt et al. \cite{Lomholt_PhysRevLett.110.208301} have enhanced our comprehension by developing a microscopic model of log-aging diffusion, which extends the Continuous Time Random Walk \cite{montroll1965random,CTRW_metzler2000random,CTRW_metzler2004restaurant} (CTRW) framework by incorporating non-renewal processes. %This model crucially utilizes the characteristics of the waiting time distribution to categorize diffusion types. 
Commonly, normal diffusion is characterized by a waiting time with a finite mean time, while subdiffusion \cite{sub_PhysRevLett.111.160604,sub_lee2021chromatin,sub_PhysRevX.6.021006} is distinguished by a waiting time distribution with a prolonged tail. Meanwhile, the timing of subsequent events in subdiffusion, influenced by the observation time \(t_a\), is modeled using the forward waiting time distribution \(\psi_1(t, t_a)\) 
\cite{forward_waiting_PhysRevLett.99.160602,forward_waiting_PhysRevLett.100.250602}. To address the enhanced memory effect, Lomholt et al. \cite{Lomholt_PhysRevLett.110.208301} adeptly adopted the waiting time distribution with functional form identical to  \(\psi_1(t, t_a)=\frac{\sin{\pi\alpha}}{\pi}\frac{t_a^\alpha}{t^\alpha(t+t_a)}\) of the subdiffusion. This conceptual advancement gives rise to ultraslow kinetics $\left<n\right> \sim \frac{1}{u(\alpha)}{\ln (t/t_0)}$, where \(\left<n\right>\) represents the ensemble average of jump steps and $t_0$ is the initial time exerting  an external perturbation, \(u(\alpha) = -\gamma - \partial\ln{\Gamma(\alpha)}/\partial{\alpha}\). This non-renewable log-aging process can be effectively demonstrated, using the analogy of a passenger's increasingly restricted mobility over time due to the subdiffusive nature of bus arrivals, as illustrated in Fig. \ref{fig1}{\color{blue}{A}}. Although these innovative correlations between jump steps and time have garnered significant attention \cite{13prl_doi:10.1126/sciadv.aao3530,13prl_PhysRevX.4.011028}, actualizing these models into measurable real-world phenomena continues to pose significant challenges.

Investigation of the target problem, involving a stationary target surrounded by one or more random walkers, is crucial in diverse disciplines, including relaxation dynamics in polymeric systems and glasses \cite{target_glass_PhysRevLett.69.478,target_glass_PhysRevLett.121.185502,target_glass__Wu2018}, complex networks \cite{network_Lau_2010,network_nature}, and molecular navigation within biological cells \cite{biology_10.1073/pnas.83.6.1608}. We derive the survival probability of the target under log-aging diffusion as a function of $\ln (t/t_0)$ (Fig. \ref{fig1}{\color{blue}{B}}), elucidating the logarithmically slow decay observed in electron glass of Anderson insulators \cite{Ovadyahu_PhysRevLett.92.066801}. Furthermore, we examine the returning probability $P(x=0,t)$, assessable via fluorescence spectroscopy, and demonstrate its dependence on time as $P(x=0,t)\sim \sqrt{\frac{u(\alpha)}{\ln (t/t_0)}}$ (see Fig. \ref{fig1}{\color{blue}{C}}). Importantly, we provide a time matching transformation between internal time and laboratory time to unify the normal, subdiffusive, and log-aging diffusion processes.

The log-aging diffusion significantly impacts system dynamics by inducing enhanced memory capabilities. The autocorrelation function emerges as an essential physical quantity for characterizing the relaxation properties of glass systems \cite{auto_glass_experiment_nature,auto_glass_PhysRevLett.77.1386,auto_glass_PhysRevLett.89.107201,auto_glass_PhysRevLett.40.589}, and has garnered substantial interest in the field of biology \cite{auto_bio_BHATTACHARYA2023105,auto_bio_doi:10.1073/pnas.1605146113}. Previous research \cite{Xie_PhysRevLett.93.180603} utilizing single-particle tracking experiments has identified that the position autocorrelation function, \(C_x = \left\langle x(t)x(0) \right\rangle\), exhibits slow decay in autocorrelation of subdiffusion described by a generalized Langevin equation. Our investigations reveal that the position autocorrelation function for the log-aging diffusion is given by $C_x \sim \left(\frac{t_0}{t}\right)^{\frac{k}{u(\alpha)}}$, with $k$ denoting a constant irrelevant to the random process, as shown in Fig. \ref{fig1}{\color{blue}{D}}, presenting a remarkable ultraslow behavior compared to subdiffusion. %This novel finding could potentially enhance our understanding of certain critical phenomena.
\section*{Logarithmic relaxation through log-aging process}
\begin{figure*}[tbhp]
\centering
\includegraphics[width=17.8cm]{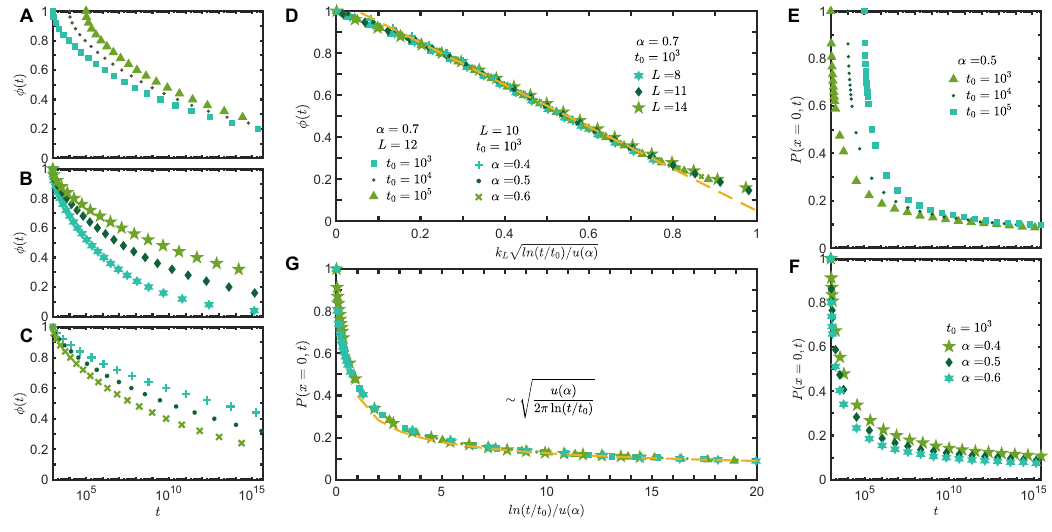}
\caption{Survival and returning probability in log-aging diffusion, manifesting universal scaling forms. (\textbf{A} to \textbf{C}), The simulated survival probability \(\phi\) of a single particle on a one-dimensional (1D) lattice is plotted, with length \(L\) and lattice spacing \(a = 1\). The particle exhibits a log-aging diffusion  process with initial time \(t_0\). 
(\textbf{D}), The survival probability data from (A to C) are appropriately scaled following Eq. \ref{eq_aging_survival}, with consistency in the symbolic representation across these sections. In panels (A to D), the same symbol denotes identical parameter values. (\textbf{E} and \textbf{F}), The simulation also explores the returning probability \(P(x = 0, t)\) for the particle initially positioned at the origin at time \(t_0\), providing insights into its likelihood to return to the starting point over time. (\textbf{G}), Data from Sections (E to F) undergo scaling as  Eq. \ref{eq_aging_P}, maintaining uniformity in symbol usage throughout the analysis. In panels (E to G), the same symbol denotes identical parameter values.
}\label{fig2}
\end{figure*}
The slow kinetics inherent to CTRW significantly influence the physical observables. A notable manifestation of this is the Williams-Watts form of relaxation, which can be derived from a CTRW characterized by a long-tail pausing-time distribution \cite{Shlesinger_PNAS_1984}. Furthermore, Montroll et al. \cite{Shlesinger_PNAS_1984} elucidated a profound link between various relaxation processes, such as current decay and dielectric function, and survival probability, the likelihood of mobile defects evading traps. Specifically, the Williams-Watts-type relaxation function correlates with the survival probability of mobile defects in 1D lattice undergoing subdiffusion as follows:
\begin{equation}
\label{eq_sub_survival}
\phi\left(t\right) \sim 1 - \frac{k_1}{L} t^{\alpha/2} \sim e^{-c \cdot k_1 \cdot t^{\alpha/2}},
\end{equation}
where \( k_1 \) is a function of \( \alpha \), \( L \) represents the length of a 1D lattice, the second expression details the survival probability of a single defect, and the final expression models the survival probability of \( N \) defects with density \( c = \frac{N}{L} \). However, despite the prevalence of the Williams-Watts type relaxation function across various physical and chemical systems \cite{William_Watts_Mongin2018,WilliamWatts_PhysRevE.99.032502,WilliamWatts_PhysRevLett.107.067801,WilliamWatts_doi:10.1073/pnas.2201566119}, exotic ultraslow logarithmic relaxation extends beyond the scope of the Williams-Watts type and necessitates a novel scheme of random processes.

The random walk characterized by log-aging diffusion, as illustrated in Fig. \ref{fig1}, inherently produces an ultraslow relaxation. We systematically calculate the survival probability across various parameters, striving for a universal scaling form. For enhanced clarity, the scenario of a single defect (N=1) is depicted in Fig. \ref{fig2}. 
The survival probability \(\phi\) as a function of time for a solitary defect undergoing log-aging diffusion in a one-dimensional system, spanning from \(0\) to \(10^{15}\) (Fig. \ref{fig2}{\color{blue}{A-C}} in a log-linear scale), unveils a distinctly ultraslow decay process. Our studies have established that the initial time \(t_{0}\), system size \(L\), and the exponent \(\alpha\) of the waiting time distribution significantly modulate the non-renewal log-aging random walk and, consequently, the survival probability. In Fig. \ref{fig2}{\color{blue}{A}}, the effect of \(t_0\) is pronounced in the short-time regime but converges over the long-time scale, indicating that the decay pattern of survival probability eventually becomes independent of \(t_0\). Fig. \ref{fig2}{\color{blue}{B}} elucidates the correlation between system size \(L\) and the rate of decay process, demonstrating that larger systems exhibit slower relaxation rates. Additionally, Fig. \ref{fig2}{\color{blue}{C}} portrays how varying the waiting time distribution exponent \(\alpha\) influences the rate of decay, revealing that lower values of \(\alpha\) correlate with a more gradual decrease in survival probability.

\begin{figure*}[tbhp]
\centering
\includegraphics[width=17.8cm]{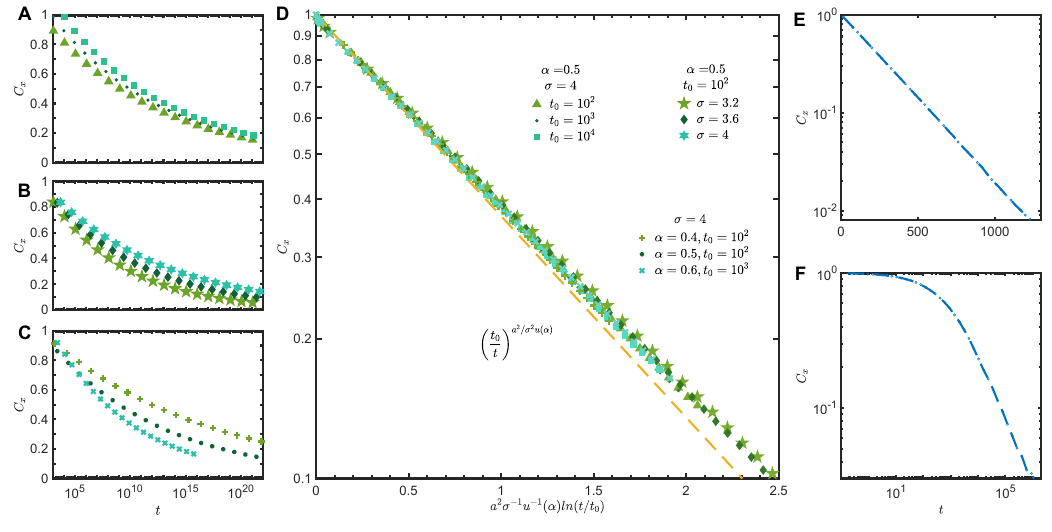}
\caption{Scheme and characteristics of log-aging diffusion.(\textbf{A}), We present a log-aging diffusion process, using a scenario where a passenger is waiting to board a bus. The intervals between bus departures at each stop are modeled by a long-tail distribution, expressed as $\psi_\alpha \sim \frac{1}{t^{1+\alpha}}$. Upon the passenger's arrival at the $n-1$ stop, they encounter a wait of $t_{n-1}$. Once the immediate bus arrives at the $n-1$ stop, the passenger boards swiftly and proceeds to the next stop $n$ with the travel time considerably shorter than the waiting time. Upon reaching stop $n$, the passenger waits for the next bus, timing the wait as $t_n$. (\textbf{B}), Displayed are the first nine states of a particle undergoing the diffusion process described in (A), parameterized by $\alpha=0.6$, $t_0=10^3$. We define $u\left(\alpha\right)=-\gamma-\partial \ln\Gamma\left(\alpha\right)/\partial \alpha$, where $\gamma$ is the Euler constant and $\Gamma$ is complete gamma function. (\textbf{C}), The return probability without external forces for the log-aging diffusion is computed as $P(x=0,t)$. (\textbf{D}), The positional autocorrelation function under the influence of an oscillator potential is given,  $C_x=\left<x(t)x(0)\right>$.}\label{fig3}
\end{figure*}

The survival probability across varied scenarios consistently exhibits ultraslow relaxation, as depicted on the log-linear scale in Fig. \ref{fig2}{\color{blue}{A-C}}. However, universal characteristics underlying these data are still desirable. Through asymptotic analysis detailed in the Supplement Materials, we have derived the asymptotic form of the survival probability for a single defect undergoing a non-renewal log-aging random walk:
\begin{equation}
\label{eq_aging_survival}
\phi\left(t\right) \sim 1 - \frac{1}{L} \sqrt{\frac{8}{\pi}} \left(\frac{\ln t/t_0}{u(\alpha)}\right)^{1/2}.
\end{equation}
Leveraging this analytical result, we configure the x-axis as: $\frac{1}{L} \sqrt{\frac{8}{\pi}} \left(\frac{\ln t/t_0}{u(\alpha)}\right)^{1/2}$, facilitating a coherent scaling form of the scattered data illustrated in Fig. \ref{fig2}{\color{blue}{D}}. Further, the survival probability for \(N\) defects, with density \(c = \frac{N}{L}\), conforms to the scaling form: $\phi\left(t\right) \sim e^{-c \sqrt{\frac{8 \ln t/t_0}{\pi u(\alpha)}}}$. Simulations for \(N\) defects are showcased in Fig. S1. This formulation of the survival probability in the log-aging diffusion provides an explanatory basis for the log-aging decay observed in glass systems \cite{Ovadyahu_PhysRevLett.92.066801} (see Supplement Materials and Fig. S2).

The ultraslow relaxation characteristic prominently influences the return probability \( P(x=0,t) \), offering profound insights into the diffusion dynamics of defects. As demonstrated in Fig. \ref{fig2}{\color{blue}{E-F}}, simulation results for log-aging diffusion in a one-dimensional infinite system, originating at the origin, reveal that \( P(x=0,t) \) decays ultraslowly over extended time periods. Although the initial time \( t_0 \) and the exponent \( \alpha \) affect the decay rate, their impact converges to consistent behavior over the long term. From our asymptotic analysis, we deduce the returning probability as follows:
\begin{equation}
\label{eq_aging_P}
P(t) \sim \frac{1}{\sqrt{2\pi u^{-1}(\alpha) \ln{t/t_0}}}.
\end{equation}
This formula indicates that the returning probability adheres to a universal scaling form, illustrated in Fig. \ref{fig2}{\color{blue}{G}}. For comparison, the asymptotic behavior of subdiffusion over extended durations is expressed by $P(t) \sim \frac{1}{\Gamma(1-\alpha/2)t^{\alpha/2}}$, highlighting a fundamental contrast with log-aging diffusion \cite{CTRW_metzler2000random}.

Upon analyzing the functional forms of the survival probability, \(\phi\), and the returning probability, \(P(x=0,t)\), we observe that both quantities for the log-aging diffusion are expressible as functions of logarithmic time, \(\ln(t/t_0)\), while for subdiffusion, they adhere to power time functions, \(t^\alpha\). These observations suggest a common underlying mechanism, governed by the number of jump steps, \(n\), that dictates the temporal evolution of the walker in internal time space, independent of the type of diffusion. In these diffusion models, the translation from internal time \(n\) to laboratory time \(t\) is significantly influenced by the waiting time distribution. The survival probability is explicitly linked to the number of steps \(n\) as \(1 - \phi \propto \sqrt{n}\) in a one-dimensional system with a single defect, and the return probability is modeled as \(P(x=0,t) \propto \frac{1}{\left<x^2\right>^{d/2}}\), with \(\left<x^2\right> \propto \left<n\right>\). By substituting \(\left<n\right> \sim \ln(t/t_0)\) for log-aging diffusion and \(\left<n\right> \sim t^\alpha\) for subdiffusion. Then, in laboratory time, more complex dynamics manifest in various random processes. Nevertheless, after integral transformation to the internal time space, the dynamics can be unified into a simple form (See Supplement Materials for the details).
\section*{Ultraslow evolution revealed by autocorrelation function}

The autocorrelation function bridges microscopic phenomena with macroscopic behavior \cite{ auto_physics_PhysRevX.4.011022}. In the time-dependent Ginzburg-Landau theory (TDGL), fluctuations result in the autocorrelation function of the order parameter that exhibits exponential decay \cite{tinkham2004introduction} . In critical dynamics, the autocorrelation function shows power-law behavior, reflecting unique dynamic properties under critical conditions \cite{tauber2014critical}.

To understand correlation in the log-aging diffusion, we investigate position-position autocorrelation \(C_x\) within a confinement potential \(U = \frac{m\omega^2 x^2}{2}\). The particle is restricted to jump only to nearest neighbor lattice sites under detailed balance conditions after initial time \(t_0\), with lattice spacing \(a=1\). The decay process of \(C_x\) is monitored up to \(t=10^{22}\), depicted in Fig. \ref{fig3}{\color{blue}{A-C}} on a log-linear scale. In particular, \(C_x\) does not approach zero even at these extensive timescales. As shown in Fig. \ref{fig3}{\color{blue}{A}}, \(C_x\) initially depends on \(t_0\), but this influence vanishes when \(t \gg t_0\). Fig. \ref{fig3}{\color{blue}{B}} shows that an increase in the variance \(\sigma\), defined as \(\sigma^2 = \frac{k_BT}{m\omega^2}\), results in a slower decay of \(C_x\). Additionally, Fig. \ref{fig3}{\color{blue}{C}} reveals that smaller values of \(\alpha\) lead to slower decay rates.

The autocorrelation function in this log-aging diffusion manifests in a scaling form:
\begin{equation}
\label{eq_aging_C_x}
C_x(t,t_0) \sim \left(\frac{t_0}{t}\right)^{a^2/\sigma^2u(\alpha)}.
\end{equation}
%As \(\sigma\) increases, thermal fluctuations dominate potential constraints, leading to the observed slow decay. Thus, 
After the scaling analysis shown in Fig. S3, one can find a remarkable feature that the autocorrelation adheres to a universal scaling form as depicted in Fig. \ref{fig3}{\color{blue}{D}}, characterized by an extremely slow power decay. 

The dynamics of log-aging diffusion processes can be modeled using a Generalized Langevin Equation (GLE) with a memory kernel: 
\begin{equation}
\label{eq_kernel_K}
K(t, t^\prime) = \frac{u(\alpha)}{\ln{t/t^\prime}}\ ,
\end{equation}
where $t>t^\prime$. The solution of this GLE results in an autocorrelation funtion that is identical to the scaling form shown in Fig. \ref{fig3}{\color{blue}{D}} (See Supplement Materials for detailed derivations). 
For normal diffusion characterized by \(K(t,t^\prime) \sim \delta(t-t^\prime)\), the autocorrelation decays exponentially (Fig. \ref{fig3}{\color{blue}{E}}). In subdiffusion, where \(K(t,t^\prime) \sim (t-t^\prime)^{-\alpha}\), \(C_x(t)\) follows a Mittag-Leffler function $C_x(t)=E_\alpha\left(-(t/\tau)^\alpha\right)$, illustrating a transition from exponential to power-law decay over time (Fig. \ref{fig3}{\color{blue}{F}}). In numerical simulations, similar random processes as shown in Fig. \ref{fig1}{\color{blue}{A}}, with various values of $0<\alpha<1$, $1\leq\alpha<2$, and $\alpha\geq2$ account for autocorrelation functions shown in Fig. \ref{fig3}{\color{blue}{D}}, Fig. \ref{fig3}{\color{blue}{F}}, Fig. \ref{fig3}{\color{blue}{E}}, respectively (see Supplement Materials for details). 

\section*{DISCUSSION}
The memory kernel of log-aging diffusion process provides a new perspective on the influences of environment on the system, from the relaxation of two-level system \cite{RevModPhys.59.1,PhysRevE.67.021111} to noise-induced anomalous diffusion \cite{PhysRevLett.114.160401,PhysRevLett.119.046601}. For example, Gaussian white noise is correlated with memoryless Markov processes in normal diffusion, while subdiffusion experiences pink noise due to significant memory effects. For log-aging processes, specific environmental noise is necessary to account for its ultraslow evolution and pronounced long-time memory. Moreover, the remarkable memory effect within the log-aging diffusion calls for in-depth investigations on the interplay of possible ergodicity breaking \cite{PhysRevLett.101.058101} and a generalized fluctuation dissipation relation.

\bibliography{ref}
\clearpage
\noindent
\includegraphics[page=1,trim=1.91cm 1cm 1cm 1.5cm,clip]{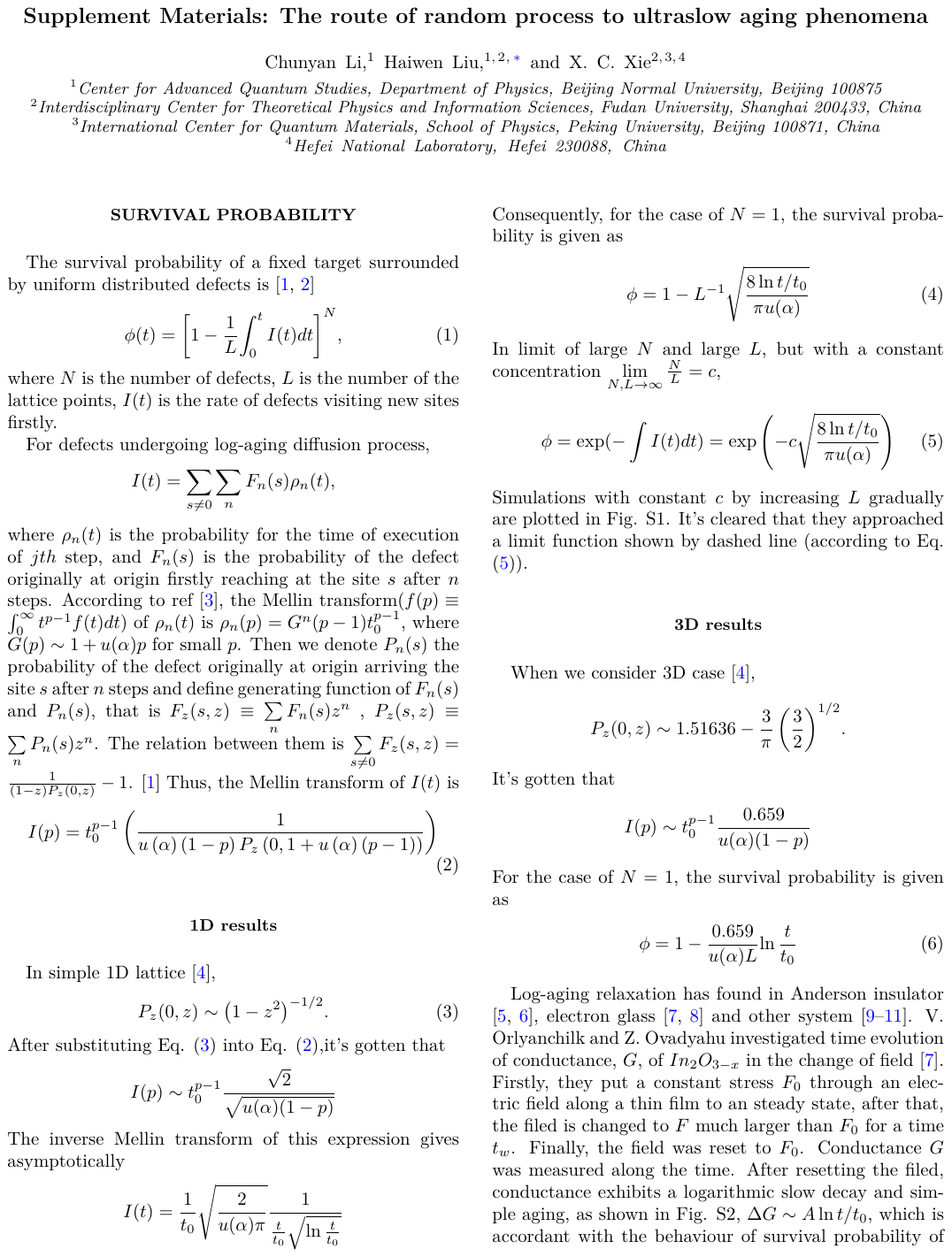}
\thispagestyle{empty}

\newpage
\noindent
\includegraphics[page=2,trim=1.91cm 1cm 1cm 1.5cm,clip]{supplementary.pdf}
\thispagestyle{empty}
\newpage
\noindent
\includegraphics[page=3,trim=1.91cm 1cm 1cm 1.5cm,clip]{supplementary.pdf}
\thispagestyle{empty}

\newpage
\noindent
\includegraphics[page=4,trim=1.91cm 1cm 1cm 1.5cm,clip]{supplementary.pdf}
\thispagestyle{empty}
\newpage
\noindent
\includegraphics[page=5,trim=1.91cm 1cm 1cm 1.5cm,clip]{supplementary.pdf}
\thispagestyle{empty}

\end{document}